\documentclass[aps,floats,twocolumn,prl,showpacs,superscriptaddress]{revtex4} 
\usepackage{graphicx}
\usepackage{amsmath}
\usepackage{subfigure}

\begin{document}

\let\n=\nu
\let\o=\omega
\let\s=\sigma
\def\np{\n'}
\def\sp{\s'}
\def\EL{E_{L}}
\def\EN{E_N}
\def\ES{E_S}
\def\EM{E_M}
\def\ExN{\mbox{e}^{-\beta \EN}}
\def\ExM{\mbox{e}^{-\beta \EM}}
\def\ExL{\mbox{e}^{-\beta \EL}}
\def\ExS{\mbox{e}^{-\beta \ES}}
\def\c{(c_{\s})}
\def\cd{(c_{\s}^{\dagger})}
\def\cp{(c_{\sp})}
\def\cpd{(c_{\sp}^{\dagger})}

%\phantom{a}

%\section{Why we hold our results to be suitable for Physical Review Letter}

%It created quite a sensation when, in 2006, kinks were reported in purely electronic systems without any bosonic degrees of freedom such as phonons or non-local spin-fluctuations. Subsequently such kinks have been observed in many different models and physical quantities.  It has been pointed out that these kinks are  mathematically necessary for lattice models with a three peak spectral function, and that they are accompanied by local spin fluctuations. However, the physical origin behind these kinds remained mysterious.

%We solve this puzzle and identify these kinks to originate from the Kondo effect of the lattice system, which -as we show-  bears quite some differences to an impurity model. While this explanation might look simple  retrospectively, we stress that up to this point this connection is not known. Since these
%kinks are a generic feature of strongly correlated lattice models, we hold our explanation not only to be insightful but also of broad interest.
%}\vfill\vfill
%\pagebreak
%\clearpage

\title{Poor Man's Understanding of Kinks Originating from Strong Electronic Correlations}

\author{K. Held}
\affiliation{Institute of Solid State Physics, Vienna University of
  Technology, A-1040 Vienna, Austria}

\author{R. Peters}
\affiliation{Department of Physics, Kyoto University, Kyoto 606-8502, Japan}

\author{A. Toschi}
\affiliation{Institute of Solid State Physics, Vienna University of
  Technology, A-1040 Vienna, Austria}

\date{\today }

\begin{abstract}
By means of dynamical mean field theory calculations,  it was recently discovered  that kinks generically arise in strongly correlated systems, even in the absence of external bosonic degrees of freedoms such as phonons. However, the physical mechanism behind these kinks remained unclear.
On the basis of the perturbative and numerical renormalization group theory, we herewith identify these kinks as the effective Kondo  energy scale of the interacting lattice system which is shown to be smaller than the width of the central peak.
\end{abstract}

% 71.27.+a  Strongly correlated electron systems; heavy fermions
% 71.10.Fd  Lattice fermion models (Hubbard model, etc.)
% 71.30.+h  Metal-insulator transitions and other electronic transitions
% 71.20.Eh  Rare earth metals and alloys
% 75.20.Hr  Local moment in compounds and alloys; Kondo effect, valence
%           fluctuations, heavy fermions
% 75.47.Gk  Colossal magnetoresistance
\maketitle

Kinks in the energy vs.\ momentum dispersion-relation indicate deviations from a  quasiparticle renormalization of the non-interacting system. Hence, these kinks provide valuable information of many-body effects. The textbook example  \cite{Ashcroft} 
 is the coupling of the electronic system  to external, bosonic degrees of freedom such as e.g.\ phonons. In this situation, a kink naturally develops at the bosonic eigenenergy. The low-energy kinks in 
high-temperature superconductors \cite{Lanzara2002,Shen2002,He2001} at 40-70 meV 
are hence taken as evidence for
an electron-phonon \cite{Lanzara2002,Shen2002} or a spin-fluctuation \cite{He2001,Hwang2004} pairing mechanism. Besides these low-energy kinks, kinks at higher energies 
have been reported, not only in cuprates \cite{Graf2006,Pan2007,Meevasana2007,Inosov2007} but
also in various transition metals \cite{Schafer2004,Menzel2005} and transition metal  oxides \cite{Yang2005,Aiura2004,Yoshida2005,Eguchi2009}. These kinks are at  50-800 meV, often  beyond the  relevant bosonic energy scales associated with phonons or non-local spin fluctuations.

On the theoretical side,  kinks at similarly high energies have been found by serendipity in local density approximation plus dynamical mean field theory (LDA+DMFT) \cite{DMFT,DMFTreview,LDADMFT1,LDADMFT2,LDADMFT3,LDADMFT4} calculations of SrVO$_3$ \cite{Nekrasov}. In these calculations the aforementioned bosonic degrees of freedom are clearly absent, and the physical origin is to be found in the strongly correlated electronic system itself. It was shown mathematically \cite{Byczuk} that a three peak spectrum with a lower and upper Hubbard band and a well pronounced central peak in-between generically results in a kink in the energy-momentum dispersion of the one-particle excitations. While it was clear, given the structure of the DMFT equations, that  the central peak of (half)width $\Gamma$ was associated with Kondo physics, the physical origin of the emergence of a second  (kink) energy scale  $\omega^*<\Gamma$ remained mysterious.
This kink also reflects in other quantities, most noteworthy the specific heat \cite{Toschi}. It has been observed as well in other materials and models such as  LaNiO$_3$ \cite{Deng12}, $f$-electron systems   \cite{Kainz12}, and
the two-band Hubbard model \cite{Greger13}.
 At the kink energy there is a maximum in the local spin susceptibility \cite{Uhrig}, which was  considered \cite{Uhrig} to represent ``emergent collective spin-fluctuations''. For two bands of different width a single maximum in the
spin susceptibility along with a single kink energy scale
has been found \cite{Greger13},  which put the generalizabilty of  \cite{Byczuk}
into question. Most of all,  a physical understanding was hitherto missing: Why
is there a second energy scale besides the width of the central peak?

\begin{figure}[t]
\includegraphics[width=8cm]{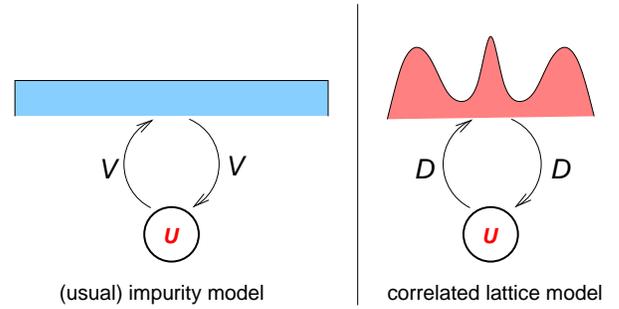}
\caption{(Color online) Comparison of the usual Anderson impurity model 
of a strongly interacting site coupled via $V$ to an uncorrelated
featureless and wide conduction electron band (left hand side) and the Hubbard model situation
(right hand side). In the latter case an electron
leaving a correlated site moves within the strongly correlated and narrow band of the central peak. In this situation  there is a kink at the effective Kondo energy scale which is smaller than the width of the narrow band.
\label{Fig:scheme}\label{Fig:ov}}
\end{figure}

In this paper, we 
identify the physical origin to be the crossover to
the strong coupling fix point.  That is, the kink corresponds to the effective 
Kondo energy scale which, for the Hubbard model, is 
different from the width of the central peak in the spectral function.
Our conclusions are based on
a very simple, albeit analytical approach, Anderson's poor man scaling \cite{Anderson} as well as numerically precise   numerical renormalization group (NRG) calculations.
In the following, we will first provide for a qualitative overview by means of Fig. \ref{Fig:scheme}. Next we present the perturbative renormalization group calculation. Thereafter,  we discuss its relevance for Hubbard-type models and transition metal oxides; and  finally the NRG results corroborating the analytical calculation.

{\em Overview.} The usual Kondo system consists of an interacting impurity site which is coupled to a non-interacting conduction band by a hybridization amplitude $V$. Usually, the conduction electron bandwidth is the largest energy scale of the system, see Fig.\ \ref{Fig:ov} (left hand side). At the Kondo energy scale a central quasiparticle peak develops. At the same energy scale the imaginary part of the susceptibility  ${\rm Im}\chi(\omega)$ exhibits a peak \cite{c5}, which can be understood as an effective scattering of quasiparticles and quasi-holes at each other. One can also consider this as a bosonic mode emerging from {\em local} spin fluctuations similar to those reported  in \cite{Uhrig}.
However, in this case, there is {\em no} kink in the  real part of the self energy,
separating two different linear behaviors.

The situation is very different if we instead consider a Kondo system with a very narrow conduction electron band, which is strongly coupled to the impurity site,  see Fig.\ \ref{Fig:ov} (right hand side). Let us stress
that this is not the usual situation considered for the Anderson impurity 
model, e.g., for describing an iron impurity in gold. However, this is the relevant situation for strongly correlated lattice models,  describing e.g.\ transition metal oxides. For such a model or material an electron leaving a site with hopping amplitude $\sim D$  enters a  strongly correlated lattice. 
Also on other lattice sites, there are hence correlation effects which lead to a renormalized, very narrow bandwidth for the central peak of the spectrum around the Fermi level. The electron considered is  moving within this very narrow band. 
 At a later time the electron might return to the original site and, possibly, interact (by local interaction $U$) with a second electron on the depicted site. 

This description of locally interacting electrons, which can  propagate via the (self-energy renormalized) other sites is at the heart of DMFT \cite{DMFTreview}; DMFT even maps the correlated lattice problem onto an Anderson impurity whose local propagator includes the described self-energy contributions from all other sites. This  Anderson impurity model is calculated self-consistently and for strong
electronic correlations has a non-interacting density of states (DOS)  as depicted in Fig.\ \ref{Fig:ov} (right hand side) \cite{DMFTreview,Moeller}.
This DMFT description neglects 
{\em non-local} correlations  such as the mentioned non-local spin fluctuations \cite{He2001,Hwang2004}. At least in three dimensions, one can however
expect DMFT to  yield  reliable results at sufficiently high temperatures or energies, such as the few hundred  meV of  high energy kinks.

As we will show below, there are two energy scales in the narrow, correlated band situation: one associated with the width of the central peak and one associated with the Kondo energy scale which is again connected  to a
 maximum in ${\rm Im} \chi(\omega)$, as well as to a stronger quasiparticle renormalization.
% \cite{footnote3}. 
This explains the observations of \cite{Uhrig} and \cite{Nekrasov,Byczuk}, respectively.
In contrast, for the usual impurity situation  considered (Fig.\ \ref{Fig:ov} left hand side) the first energy scale, i.e., the bandwidth of the central peak is missing, since the conduction electron bandwidth is essentially infinite. Here,  only the Kondo energy scale remains.

%We can hence, as is depicted in the upper panel of Fig.\ \ref{Fig:ov},
%attribute the  
%In other words, and this is 

{\em Poor man's scaling}. In Anderson's perturbative renormalization group, the conduction electrons are eliminated step-by-step by reducing the bandwidth of the conduction electrons from $[-{\cal D},{\cal D}]$ to  $[-({\cal D}-{\rm d}{\cal D}),({\cal D}-{\rm d}{\cal D})]$ in the Kondo model \cite{Anderson,commentKM}. This renormalizes the interaction $J=4V^2/U$ between impurity spin and conduction spin by \cite{Anderson,Hewson} 
\begin{equation}
{{\rm d} J({\cal D})}/{{\rm d} {\rm ln} {\cal D} }= -2 \rho({\cal D}) J^2({\cal D}) \,. \label{Eq:pRG}
\end{equation}
Here,  $\rho({\cal D})$ is the DOS of the  conduction electrons at the energy ${\cal D}$ and $-{\cal D}$ around which the conduction electrons are integrated out by second order perturbation theory.

Usually, $\rho({\cal D})$ is taken constant which  results in a Kondo temperature
$T_K={\cal D}_0 e^{-1/(\rho_0 J({\cal D}_0))}$ \cite{Anderson,Hewson}. 
In our case,
a constant density of states  is, however, certainly not appropriate. Hence, we now employ  Anderson's
poor man scaling for the situation depicted in  Fig.\ \ref{Fig:ov} (right hand side) instead of the constant one (left hand side). A reasonable description for the conduction electron DOS arising from strong correlations is a Lorentzian
 $\rho({\cal D})=\rho_0 \Gamma^2/({\cal D}^2+\Gamma^2)$ of width $\Gamma$, the 
width of the central spectral peak. In this case the integration of Eq. (\ref{Eq:pRG}) from the initial band edge ${\cal D}_0$ to ${\cal D}$ yields 
\begin{equation}
 \frac{1}{J({\cal D})}- \frac{1}{J({\cal D}_0)} = \rho_0  {\rm ln}\big( \frac{{\cal D}^2}{{\cal D}^2+\Gamma^2}\big) \big|_{{\cal D}_0}^{\cal D} .
\end{equation}
Collecting all terms with  cutoff ${\cal D} $ and ${\cal D}_0$ on the left and right hand side, respectively, yields,
\begin{eqnarray}
 \frac{{\cal D}^2}{{\cal D}^2 +  \Gamma^2} e^{-{1}/(J({\cal D})\rho_0)}&=& \frac{{\cal D}_0^2}{{\cal D}_0^2+\Gamma^2} e^{-{1}/(J({\cal D}_0)\rho_0)} \;\; \\
&\stackrel{{\cal D}_0\rightarrow \infty}{\longrightarrow}& e^{-{1}/(J({\cal D}_0)\rho_0)}= const.  \;\;\label{Eq:Tknew}
\end{eqnarray}
%\begin{equation}
% \frac{{\cal D}^2}{{\cal D}^2 \!+ \! \Gamma^2} e^{\frac{-{1}}{J({\cal D})\rho_0}}\!=\! \frac{{\cal D}_0^2}{{\cal D}_0^2\!+\!\Gamma^2} e^{\frac{-{1}}{J({\cal D}_0\rho_0)}}
%\stackrel{{\cal D}_0\rightarrow \infty}{\longrightarrow}  e^{\frac{-{1}}{J({\cal D}_0\rho_0)}} \!=\! const.  \;\;\label{Eq:Tknew}
%\end{equation}

Here, we have set the initial cut-off ${\cal D}_0$ to infinity in the second line, and identified the combination of ${\cal D}$ and $J({\cal D})$ on the left hand side to be invariant under the renormalization group flow.
This can be compared to the usual poor man's scaling result \cite{Anderson,Hewson} for a constant DOS, i.e., $\rho({\cal D})=\rho_0$:
\begin{equation}
{\cal D} e^{-{1}/(J({\cal D})\rho_0)}={\cal D}_0 e^{-{1}/({J({\cal D}_0)\rho_0})}= const. \equiv T_K \;.
\label{Eq:Tk}
\end{equation}
If the energy (cutoff) ${\cal D}$ approaches the Kondo temperature in Eq. (\ref{Eq:Tk}), the coupling $J({\cal D})$ diverges. This marks the crossover to the  strong coupling fix point in the renormalization group flow.

For the narrow conduction band case Eq.\ (\ref{Eq:Tknew}) on the other hand, this divergence of $J$ and hence the strong coupling fix point is approached for
\begin{equation}
{\cal D}= \omega^*=\sqrt{{\eta}/({1-\eta})} \; \Gamma \;\;\; \text{ with } \eta=e^{-1/(J({\cal D}_0)\rho_0)} \;.
\label{eq:omegaK}
\end{equation}
That is besides the conduction electron (half) bandwidth $\Gamma$, there is a second energy scale $\omega^*$ in the problem, at which the Kondo effect marks the 
crossover to the strong coupling fix point. This crossover is
 accompanied by strong local spin fluctuations, connected to the above mentioned maximum  in ${\rm Im \chi}$ \cite{c5}, and the stronger quasiparticle renormalization
of the strong coupling fix point.

{\em Relevance for DMFT and kinks in transition metal oxides.}
The two energy scales $\Gamma$ and $\omega^*$ are relevant for strongly correlated electron systems with a central peak. For the one-band Hubbard model with semicircular  DOS (Bethe lattice)
and half bandwidth $D$, the
  DMFT self-consistent Anderson impurity model has the following non-interacting Green function \cite{DMFTreview}:
\begin{equation}
{\cal G}_0^{-1}(\omega) = \omega + \mu -(D/2)^2 \; G(\omega).
\label{Eq:Bethe}
\end{equation}
 For a general DOS, there are corrections to Eq.\ (\ref{Eq:Bethe}), which  however still remains the leading term in a momentum expansion of the DOS.

%\begin{figure*}[tb]
\begin{figure}[tb]
%\begin{minipage}{.69 \textwidth}
%\includegraphics[width=8.4cm]{newplot.eps}
\includegraphics[width=8.4cm]{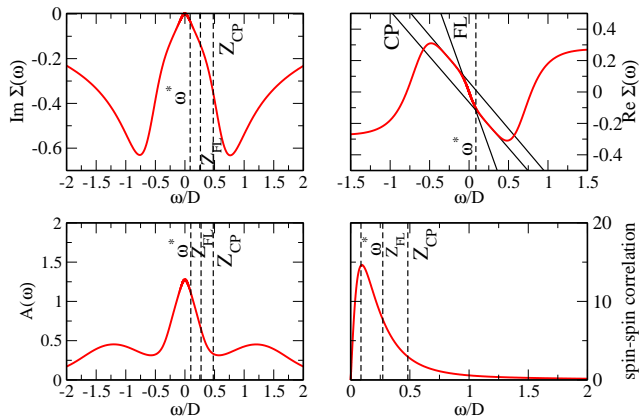}
%%\end{minipage}\hfill
%\begin{minipage}{.3 \textwidth}
\caption{(Color online) DMFT(NRG) results for the Hubbard model at $U=2D$.
Upper panels: imaginary (left) and real part (right)
 of the self energy. The latter shows a kink at $\omega^*$; the linear 
slopes before and after  $\omega^*$ (straight lines) define a Fermi liquid  ($Z_{\rm FL}$)  and central peak renormalization factor  ($Z_{\rm CP}$), respectively, whose values
are indicated in the other panels.
Lower left panel:  spectral function $A(\omega)$.
Lower right panel: spin-spin correlation function with a maximum at $\omega^*$.
\label{Fig:NRG1}}
%\end{minipage}
\end{figure}

At the same time, this non-interacting Green function 
${\cal G}_0$  is connected to the hybridization $V$ and the non-interacting conduction-electron Green function $G_0$ of the Anderson impurity model through
\begin{equation}
{\cal G}_0^{-1}(\omega) = \omega + \mu -V^2 G_0(\omega).
\label{Eq:AIM}
\end{equation}
As already depicted in Fig.\   \ref{Fig:scheme} (right hand side), this
local non-interacting Green function ${\cal G}_0$ stems from hopping processes, with electrons leaving the impurity site with amplitude  $D$ ($V$), moving through a narrow conduction band with DOS $\rho(\omega)=-\frac{1}{\pi} {\rm Im}G_0(\omega)
=-\frac{1}{\pi} {\rm Im}G(\omega)$ [if we take       $V=D/2$, note that only the combination  $\rho(\omega)V^2$ is relevant in Eq.\ (\ref{Eq:AIM})].

 We can disregard the
Hubbard side bands of DMFT in ${\rm Im}G(\omega)$ or  $\rho(\omega)$ since virtual excitations at large energies are suppressed in the renormalization group flow (only yield a negligible renormalization of $J$). Therefore, we can concentrate on the central peak whose 
spectral function $A(\omega)=-\frac{1}{\pi}{\rm Im}G(\omega)$  can be approximated by a Lorentzian of width $\Gamma$ and height $\rho(0)=2/(\pi D)$. The latter is  pinned to its non-interacting value \cite{DMFTreview}.

For half-filling and a narrow enough central peak, we are in the Kondo regime so that we can map the  Anderson impurity model directly onto a Kondo model with $J=4V^2/U$.
In other cases, this is also possible but only after first renormalizing the parameters of the Anderson impurity model itself \cite{Hewson}.
For this $J$ and a typical value of  $U=2D$ for a three peak spectrum, we obtain $\omega^*=0.21 \Gamma$ from Eq.\ (\ref{eq:omegaK}); for a larger value of $U=2.8D$ we obtain   $\omega^*=0.11 \Gamma$. Hence, the Kondo and kink energy scale
  $\omega^*$ is directly related to the (half)width of the central peak  $\Gamma$;
and both of them get smaller and smaller when we approach the Mott-Hubbard transition.
Note, also $Z_{\rm FL} D$ is directly related to $\Gamma$ (or  $Z_{\rm CP}D$), see \cite{Moeller,Bulla99}.

What do we have in the energy region  $[\omega^*,\Gamma]$ if
the Kondo effect only sets in at  $\omega^*$?
Here, in the DMFT the parameters are such that  $J$ and $\rho(\omega)$ are large
even without a renormalization of $J$ as soon as $\omega\lesssim\Gamma$. Hence, even without the Kondo effect, there is already  
spectral weight in the central peak  for  $\omega\in[\omega^*,\Gamma]$.
At   $\omega^*$, the Kondo effect then strongly renormalizes $J$,
which translates into a much stronger renormalization of the quasiparticles
and a kink in the self energy.

Indeed, in DMFT we have necessarily  $\Gamma>\omega^*$. 
Otherwise, i.e., for $\omega^*=\Gamma$,
the renormalization group flow from an {\em infinitesimally}
small energy interval around $\omega^*=\Gamma$
would strongly renormalize  $J$ to the strong coupling fix point, which
is mathematically not possible. 
%In the
% energy interval $[\omega^*,\Gamma]$, the initial $J$ is large and
%it gets noticeably renormalized, maybe comparable to the behavior at energies just above
%the Kondo energy scale in the usual Anderson impurity model.
%If $A(\omega)$ has a three peak structure in the DMFT, this
% ends however rather abruptly since $\rho(\omega)=A(\omega)$
%vanishes (or at least becomes small) at
%$\Gamma$. 
The bandwidth of the central peak
 hence defines the second energy scale $\Gamma>\omega^*$.
While the Kondo energy is $\omega^*$, the Kondo effect indirectly generates
also the energy scale $\Gamma$  through the DMFT self-consistency, which physically describes that there is similar Kondo physics on the neighboring sites. There is a
strongly enhanced coupling even above the Kondo scale $\omega^*$ but not beyond $\Gamma$.

{\em Numerical renormalization group.}
It is well known  \cite{Hewson} that terms in 3rd order perturbation theory and beyond may modify  the Kondo temperature. Hence, we have also employed the numerical renormalization group (NRG) \cite{NRGref,Krish,Bullarev} with cutoff parameter $\Lambda=2$.
Fig.\ \ref{Fig:NRG1} shows the DMFT(NRG) self energy, spectral function
and  spin susceptibility for the Hubbard model at $U=2D$ with Bethe DOS. 
Clearly, there is a kink at $\omega^*$ in the real part of the self energy. The slopes of the self energy before and after
the kink define two different renormalization factors  
%\begin{equation}
$
Z_{\rm FL (CP)}= \big[1-\partial
\text{Re}  \Sigma(\omega)/\partial \omega\big|_{\omega<\omega^*(\omega>\omega^*)}\big]^{-1}
$
%\end{equation} 
with $Z_{\rm FL} < Z_{\rm CP}$.
The overall halfwidth of the central peak is
$\Gamma=Z_{\rm CP} D$ so that we can read of the kink energy in Fig.\ \ref{Fig:NRG1} as $\omega^*\sim 0.21 \Gamma$, 
in agreement with the poor man's scaling prediction. The same holds for
$U=2.8D$, where NRG yields  $\omega^*=0.004D$ and $\Gamma=0.036 D$, i.e.,
 $\omega^* \sim 0.11  \Gamma$ in unexpectedly good agreement with poor man's scaling.
At the kink energy  $\omega^* < \Gamma$, 
the spin susceptibility in the lower left panel of Fig.\ \ref{Fig:NRG1}
has a maximum.

\begin{figure}[tb]
%\begin{minipage}{.65 \textwidth}
%\begin{minipage}{.65 \textwidth}
\includegraphics[width=5.5cm]{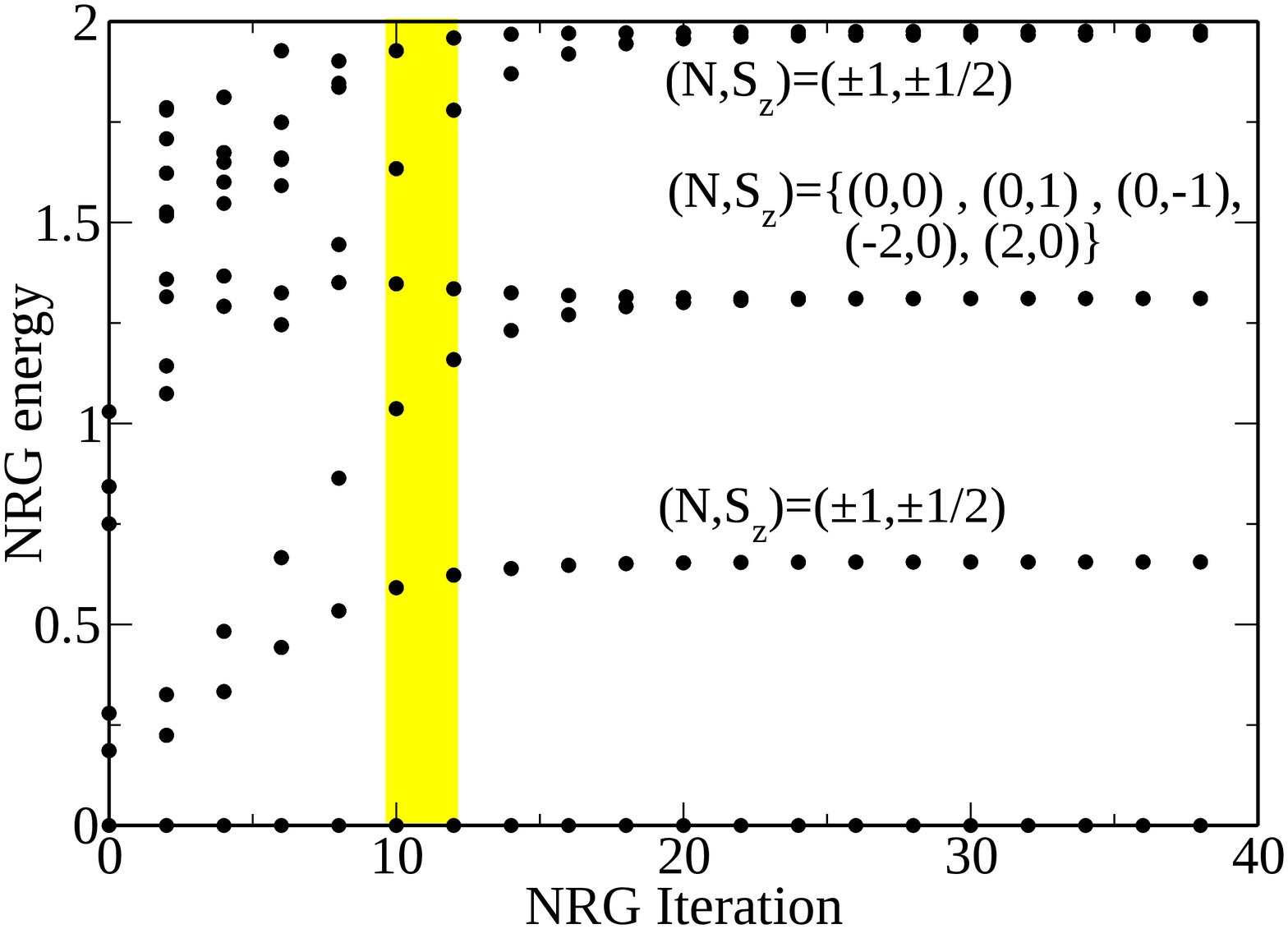} \\
\vspace{.4cm}

\includegraphics[width=5.5cm]{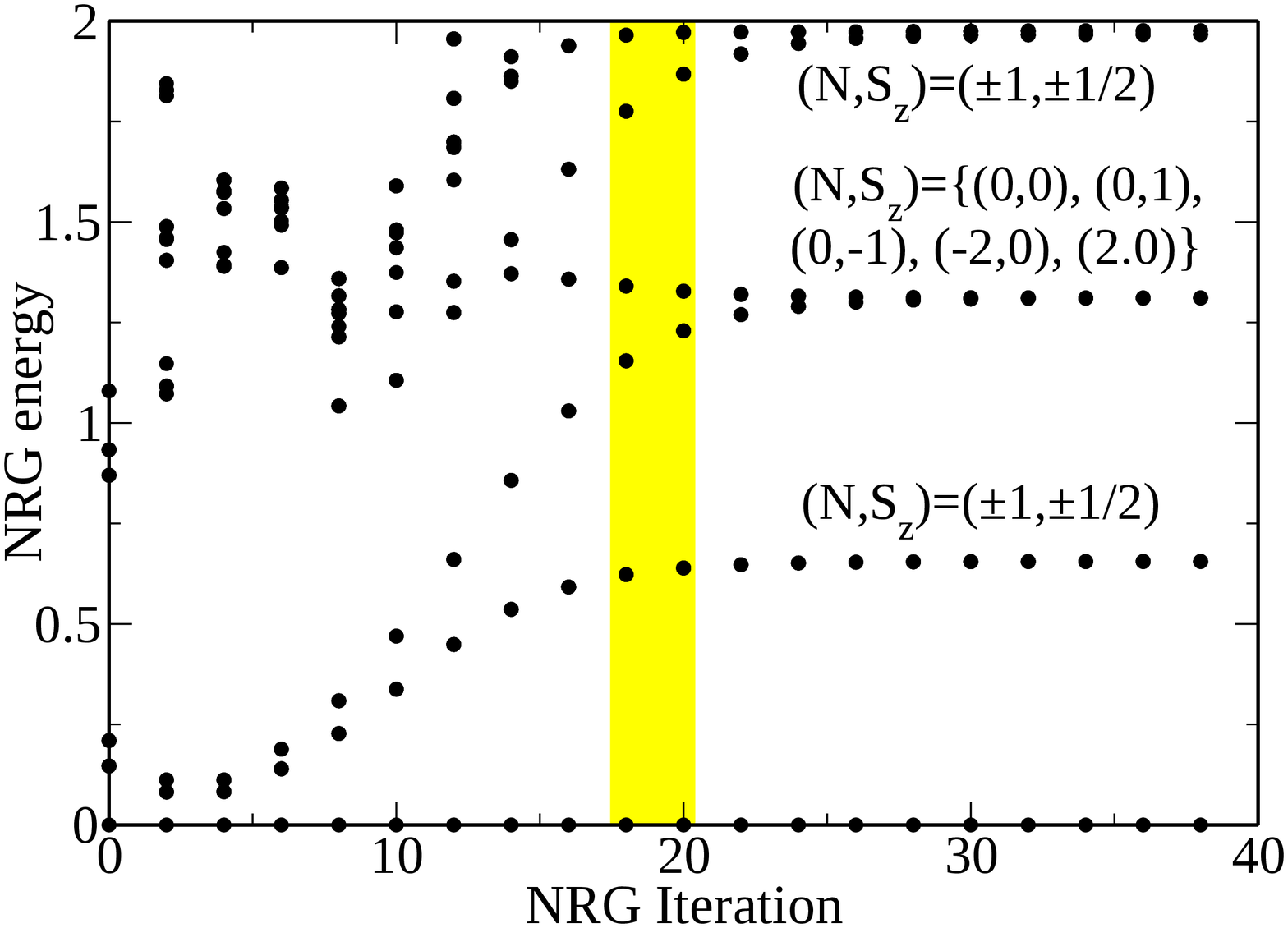}
%\end{minipage}\hfill
%\begin{minipage}{.34 \textwidth}
\caption{(Color online) DMFT(NRG) energy levels vs.\  NRG iteration
% (the upper x-axis shows the corresponding energy)
 for the Hubbard model at $U=2D$ (upper panel) and $U=2.8D$ (lower panel).
The quantum numbers in brackets 
indicate the difference in charge  ($N$) and  spin-$S_z$ component 
compared to the lowest energy level whose energy we fixed to zero.
The kink energy $\omega^*$ corresponds to the NRG iterations  highlighted in yellow/grey-shaded.
 The NRG energy levels indicate the crossover to the strong coupling fix point at $\omega^*$.\label{Fig:NRG2}}
%\end{minipage}
\end{figure}

To further elucidate that $\omega^*$ indeed represents the
crossover to the strong coupling fix point, we present in
Fig.\ \ref{Fig:NRG2} the lowest NRG energy levels as a function of the NRG iteration. At iteration $i$, energies 
$\omega=1/2 (1+1/\Lambda)*\Lambda^{-(i-2)/2}D$ become accessible within the NRG flow, and the iterations highlighted in
Fig.\ \ref{Fig:NRG2} are those were  the
kink energy $\omega^*$ is reached.
In this region, the NRG energy levels show a crossover to the strong
coupling fix point. After this crossover, the energy levels remain constant, i.e.,
at the strong coupling fix point,
for subsequent iterations. 

At $U=2.8D$ (lower panel of Fig.\ \ref{Fig:NRG2}),
there is
a considerable rearrangement of the energy levels around iteration 2-8. This might possibly correspond to a crossover from the 
free orbital to the local moment fix point of the Anderson impurity model.
However since this is restricted to a few iterations, there is no clear local moment plateau as in \cite{Krish}, where a larger ratio $U/(V^2 \rho_0)$ has been employed.
% \cite{footnote1}.  
%On physical grounds, we would also expect such a  local moment fix point energy window. This is because 
%kinks only occur if the DMFT spectral function has three peaks with a minimum between
%central peak and Hubbard bands  \cite{Byczuk}. And the physics at the energy scale of these  Hubbard bands is that of local moments.
At larger iterations,
which correspond to the kink energy  $\omega^*$ at  $U=2.8D$,
we see again  the final crossover to the  strong coupling fix point.

{\em Conclusion.} Using  poor man's scaling, we have shown that 
the kink energy  $\omega^*$ is actually the Kondo energy scale which is smaller
than  the (half)width $\Gamma$ of the central peak of a strongly correlated electron system.
At $\omega^*$,  we find the crossover to the strong coupling 
fix point  which enhances the coupling strength and with this the quasiparticle renormalization. Hence, there is  a kink in the self energy. 
Let us emphasize that this is a radically new insight; the present-day DMFT
understanding is that the Kondo effect sets in already at $\Gamma$.
The crossover to the strong coupling fix point naturally leads to a maximum in the local spin susceptibility at  $\omega^*$ as was reported in \cite{Uhrig,Greger13}. 
The same maximal spin susceptibility is also found at the Kondo
energy scale of  the usual Anderson impurity model with
a wide conduction electron bandwidth \cite{c5}. However, in the latter case, there
is no kink since the Kondo energy scale is the only low energy scale.
In a two orbital model, there will be typically a joint SU(4) Kondo effect of all orbitals, which explains the single kink enegry found in  \cite{Greger13}.
% We have also investigated the NRG energy levels which corroborates the crossover to the strong coupling fix point
%at the kink energy $\omega^*$.  
% Hence, we conclude that the Kondo effect, which has been so essential for
%our understanding of spin-carrying impurities and quantum dots, is also responsible 
%for kinks in lattice models.

%, in the unusual setup of a narrow band Anderson impurity model, naturally explains the kinks in the self energy
%and the maximum in the spin susceptibility. 

This explanation allows for distinguishing this kink from other
kinks of different origin by searching in experiment for the typical 
Kondo physics  \cite{Hewson} (keeping in mind
the additional physics  emerging between $\omega^*$ and $\Gamma$).
If one observes a kink in the energy-momentum relation
of angular resolved photoemission spectroscopy (ARPES), the origin as a Kondo kink
will be demonstrated by a simultaneous observation of a maximum in the frequency or temperature dependence of the susceptibility, the temperature dependence of the nuclear magnetic resonance (NMR) $T_1$ relaxation time, a change of the $T^2$ behavior 
in the resistivity, and a kink in the electronic specific heat.

{\em Acknowledgements.}
This work has been supported by the European Research Council under
the European Union's Seventh Framework Programme (FP/2007-2013)/ERC through grant agreement n.\ 306447 (KH), the Japan Society for the Promotion of Science (JSPS) through the FIRST Program (RP), and
 the Austrian Science Fund (FWF) via Research Unit FOR
1346 (AT, FWF project ID  I597).


\begin{thebibliography}{10}
\bibitem{commentKM} If one takes the Anderson impurity model, one possibly needs a second renormalizion before reaching the regime where the mapping the Kondo
model is feasible \cite{Hewson}.
\end{thebibliography}
\end{document}